# *"Heat Flowing from Cold to Hot without External Intervention"* Demystified: Thermal-Transformer and Temperature Oscillator

Milivoje M. Kostic

*Professor Emeritus of Mechanical Engineering*
Northern Illinois University, USA
Email: kostic@niu.edu ; Web: www.kostic.niu.edu

**A recent *Science Advances* paper by Schilling *et al*, claiming "flow of heat from cold to hot without intervention" with "oscillatory thermal inertia" are fundamentally misplaced and dramatized as miraculous, even though compliance with the Second Law of thermodynamics is acknowledged. There is nothing "magical and beyond the proof-of-concept" as claimed. It could have been achieved by any work generating device, stored by any suitable device (superconductive inductor was beneficial but not essential as claimed), and such stored work used subsequently in any refrigeration device to sub-cool the body. Cooling devices work by transforming temperature to desired level by work transfer (thermal transformer and temperature oscillator), by non-thermal, adiabatic processes. However, the "direct heat transfer" is always from higher to lower temperature in all refrigeration components, without exception – it is not to be confused by "net-transport of thermal energy by work" from cold to hot ambients. The unjustified claims are critically analyzed and demystified here.**

## INTRODUCTION

This writing is instigated by the recent publication, "Heat flowing from cold to hot without external intervention by using a 'thermal inductor'" (1), as well as by other similar, subtle and elusive thermal phenomena. The authors, regardless of acknowledging non-violation of the Second Law, have interpreted their results as a "thermodynamic magic" and made fundamentally-false critical claims. However, it was only an interesting and creative "show-and-tell" innovative application, although impractical, but not miraculous nor unusual and "beyond the-proof-of-concept" as presented. Being published in a reputable journal (1) and along with dramatic news releases (2), followed by intriguing popular articles, "*Bending (But not Breaking) the Second Law*" (3) and others, it resulted in misleading and unusually widespread publicity, even hype. There is a need to dissect and rectify the published misleading claims and demystify all magic by presenting due critical reasoning and subtle thermodynamic analysis related to above publication and ever-increasing challenges to the Second Law of thermodynamics (4), which is the main objective of this article.

There is nothing wrong with the calculated results using simplified modeling nor with elaborative but very inefficient experimental results, with barely visible outcome of only about 2 ºC sub-cooling and negligible efficiency, regardless of use of superconductive inductor (1). However, the fundamentally false claims and dramatic interpretation of the results published in a reputable journal, require objective, phenomenological reasoning and due thermodynamic analysis, beyond the mathematical modeling. The following, fundamentally false claims: (1.) "*Heat flow from cold to hot without external intervention, and without any source of power*," (2.) "*Direct heat flow from cold to hot in the Peltier TE element*," (3.) "*Use of inductor is essential*," and (4.) "*Some kind of 'thermal inertia'*," are demystified in full details here.

As a matter of fact, it would be violation of the Second Law of thermodynamics for "heat to [actually] directly flow from cold to hot" and it never happens in nature, it would be physically impossible to destroy entropy. Without exception, including all classical refrigeration and other cooling devices and their components (evaporators, condensers, etc.) the heat is always, actually transferred from hot to cold, and never otherwise. This is not to be confused with the net-transport of heat from cold to hot ambients, transported with material medium (an electron stream or refrigerant in general) while temperature of the medium is transformed by work, increased or decreased to drive heat transfer as desired (as in all thermal-power generation, cooling or heat pump devices). The Fourier's, direct heat transfer (from hot to cold only) should not be confused with transport of thermal energy (from any to any temperature level). The heat is actually transferred from the subcooled body to the even lower temperature of the Peltier cold end-plate (as is done from the cold food in a refrigerator to even colder evaporator plate inside), and after being adiabatically transported with temperature increase by electron stream (similar to a refrigerant medium transport and compression in a refrigerator), the heat is then transferred from higher temperature of the Peltier hot end-plate to the lower temperature of a surrounding reservoir (as is done from the hot condenser fins to the colder ambient air outside a refrigerator) -- the heat being always, actually transferred from hot to cold, and never otherwise, see Fig. 1.

In fact, since the heat could be net-transferred, i.e., transported from any-to-any temperature level using relevant thermal cycles and related devices (e.g., heat engines, other thermal-work generators, refrigerators and heat pumps), they may be considered as "thermal transformers," where the temperature level could be transformed to any desired level, like the voltage is in electrical transformers.

There is a need to further illuminate and demystify thermodynamic facts from confusion and mystique, since there are many puzzling issues still surrounding thermodynamics and the nature of heat, including subtle definitions and ambiguous meanings of some very fundamental concepts (5, 6).

The nature of heat was intriguing since its introduction, starting from Lavoisier, who presumed that caloric as a weightless substance is conserved, to Sadi Carnot (7) who erroneously assumed that work is extracted while caloric is conserved, to modern day researchers who argue that thermal energy is an indistinguishable part of internal energy, to the generalization of entropy and challengers of the Second Law of thermodynamics (4, 8). In fact, all other forms of energies are ultimately dissipated in thermal heat, the omnipresent and universal phenomena, quantified with perpetual and irreversible generation of entropy, i.e., the thermal displacement (9). Sometimes, highly accomplished scientists in their fields and even some thermodynamicists, do not fully comprehend the essence of the Second law of thermodynamics, and misrepresent the elusive thermal phenomena, let alone the surrounding publicity and even hype, resulting in speculative challenges of the Second Law of thermodynamics (4).

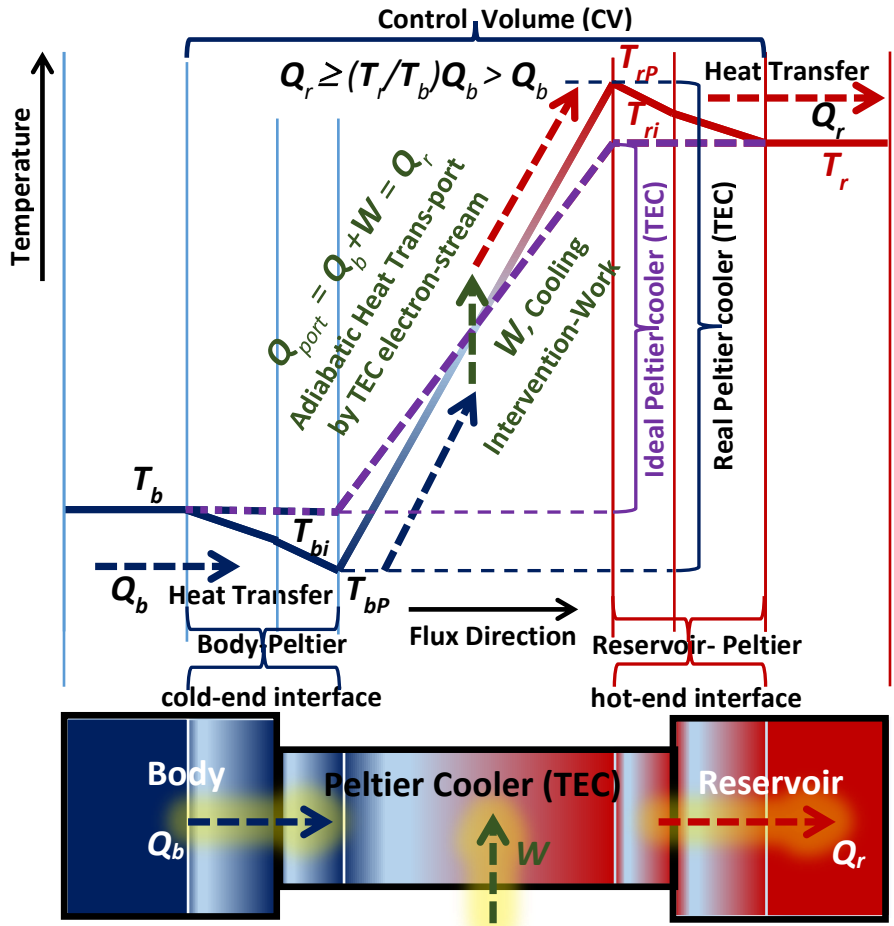

**Fig. 1. Temperature-flux diagram of Peltier thermo-electric cooler**, **TEC**, (simplified, not in scale). The system control-volume (CV) includes TEC with cold- and hot-end interfaces. The cooling-load heat ($\dot{Q}_b$), transported by TEC electron stream ($\dot{Q}_{port}$) from the body (*b*), through the Peltier thermo-electric cooler (TEC), into the surrounding reservoir ($\dot{Q}_r$). During the cooling process the electron stream temperature is decreased below $T_b$ to $T_{bP}$ at the body- Peltier, cold-end interface, and increased above $T_r$ to $T_{rP}$ at the reservoir-Peltier, hot-end interface, so that actual heat is being always spontaneously transferred from higher to lower temperature locally, while overall transported (using cooling intervention-work $W$), from cold-end to hot-end of Peltier TEC, like in all other refrigeration devices (adiabatic "heat transport" is different from actual heat transfer). The "direct heat transfer" (as hypothesized in Ref.1) is not a real heat-transfer but actually, adiabatic heat transport ($\dot{Q}_{port} = \dot{Q}_b + \dot{W} = \dot{Q}_r$), carried by a flow of a coolant medium (the Peltier electron stream or refrigerant medium in general). The adiabatic heat transport is not associated with entropy and ideally does not generate any entropy.

The authors of "*Heat flowing from cold to hot ...*" (1) present a very creative physical application, albeit inefficient and impractical, to self-sub-cool a hot body relative to intermediate ambient temperature using its own initial, thermal work potential stored within. However, the results have been presented in somewhat spectacular and misleading ways, including dramatic news releases followed by wide publicity in a number of media, as evidenced by a Google search titled "Thermodynamic magic enables cooling without energy consumption," or "Heat flowing from cold to hot without external intervention." The following authors' statements instigated the follow up wide publicity bordering with magics: *"Intriguingly, the process initially appears to contradict the fundamental laws of physics ... appear at first sight to challenge the second law of thermodynamics ... Theoretically, this experimental device could turn boiling water to ice, without using any energy ... With this very simple technology, large amounts of hot solid, liquid or gaseous materials could be cooled to well below room temperature without any energy consumption ... At first sight, the experiments appear to be a kind of thermodynamic magic, thereby challenging to some extent our traditional perceptions of the flow of heat… etc.".* The unusually widespread media coverage, implying magical phenomena, is mostly due to the lack of full comprehension of the still elusive Second Law of thermodynamics and subtle issues related to thermal phenomena.

**SELF COOLING OF HOT BODY AND CARNOT CYCLE LIMITATIONS**

An innovative "proof-of-concept" physical application, analyzed and experimentally tested in (1), is revisited here again using ideal, reversible processes, depicted on dimensionless temperature-entropy diagram on Fig. 2, and minimum sub-cooling temperature calculated for different irreversible efficiencies are presented in Table 1, using the same data and nomenclature, for convenient comparison and analysis. Note that subscript "*b*" for copper-body in (1) is dropped in here for simplicity, for all "body" quantities, i.e., $T_b \equiv T$, and all correlations are presented for a unit mass of the body.

**The cooling (Stage 1) and sub-cooling (Stage 2) of a hot body**

On Fig. 2 the relevant dimensionless temperature-entropy, $\theta$ -$\sigma$ (corresponding to *T-S*) diagram, is depicted for copper body cooled from $\theta_0$ to $\theta_r$ (Stage 1) by ideal heat-work generator (G) (like ideal Peltier generator-element or ideal heat engine cycle R-O-P-R) connected to an ideal work storage and retrieval device (S) (like ideal inductor, or capacitor, or spring, or elevated mass), and then sub-cooled in Stage 2, below ambient temperature, from $\theta_r$ to $\theta_{min}$ by ideal cooling device (C) (like ideal Peltier cooler or ideal refrigeration cycle M-R-N-M). The ideal heat-work-refrigeration devices are analyzed as the ideal Carnot cycles (ideal thermal transformers) working with infinitesimally small energy transfers, $\delta W = \eta_C \delta Q$, and integrated over the relevant variable temperature ranges or related entropy ranges. Furthermore, for $C_v=0.385\ kJ/K/kg$, a constant specific heat of a copper body, the temperature will be an exponential function of entropy, also dimensionless temperature, $\theta = T/T_r$, and dimensionless entropy, $\sigma = (S-S_r)/C_v$ (see Fig. 2):

$$S = \int \frac{C_v dT}{T} = C_v \cdot ln\left(\frac{T}{T_r}\right) + S_r \text{ or,} \tag{1}$$

$$\frac{T}{T_r} = exp\left(\frac{S-S_r}{C_v}\right) \text{ or } \theta = e^\sigma \tag{2}$$

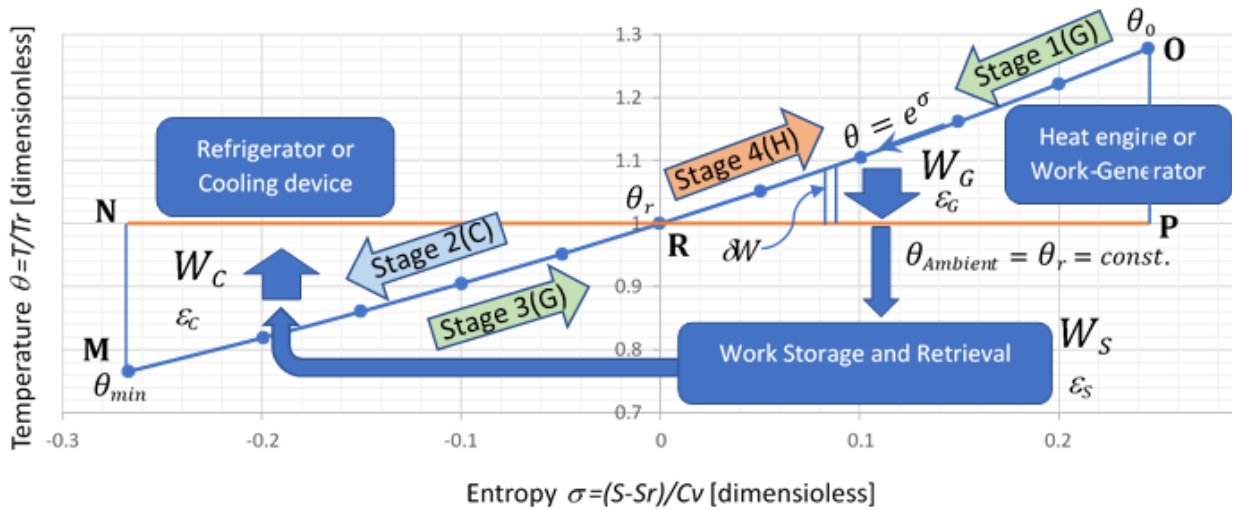

**Fig 2. Self-cooling (Stage 1) and sub-cooling (Stage 2) of a hot body using thermal work-generator (G), storage (S) and cooling (C) devices.** For 104 °C initial body temperature ($\theta_0$=1.28), the maximum sub-cooling with ideal, reversible devices within 22 °C reference ambient ($T_r$=295 K), the -47 °C minimum temperature ($\theta_{min}$ =0.766) would be achieved. If efficiencies of work generator, storage, retrieval and cooler are 0.6*0.9*0.9*0.6 respectively (total about 30%, as in classical efficient heat engine, elevated mass-storage-retrieval and refrigerator devices), negative -17.4 °C, minimum body temperature would be achieved. If thermo-electrical Peltier element with capacitive storage is used resulting in about 0.5% total irreversible efficiency (0.1*0.1*0.75*0.75=0.005) or about 5 °C subcooling or 16.7 °C minimum body temperature would be achieved [see Eq.(7) and Table 1 for more]. NOTE that when the body is approaching ambient temperature the work is generated, Stages 1(G) & 3(G), cycles ROPR and RMNR, respectively, while when going away from ambient, Stage 2(C) sub-Cooling (cycle MRNM) or Stage 4(H) Heat-pump heating (cycle RPOR), prior stored work is used.

The work of an infinitesimal Carnot cycle powered by the hot body of variable $T$ with constant heat capacity $C_v$ within an ambient reservoir at $T_r$, is:

$$\delta W = \eta_C \delta Q = \frac{T-T_r}{T} \underbrace{\delta Q}_{C_v dT}^{TdS} = (T - T_r)dS = \left(1 - \frac{T_r}{T}\right) \cdot C_v dT \qquad (3)$$

The maximum possible work to be extracted while running an ideal heat engine or ideal thermo-electric (TE) generator (G), by cooling the hot copper body from $T_0$=377 K =104 °C to $T_r$=295 K =22 °C ambient reference temperature, where dimensionless temperature $\theta=T/T_r$ and dimensionless entropy $\sigma=(S-S_r)/C_v$, is:

$$W_G = -C_v T_r \int_{\theta_0}^{\theta_r} \left(1 - \frac{1}{\theta}\right) d\theta = C_v T_r [\theta_0 - 1 - \ln(\theta_0)] \qquad (4)$$

And for incompressible systems with constant specific heat, having the variable temperature as the heat source, $\theta = exp(\sigma)$, the Carnot efficiency is:

$$\eta_C = \frac{W_G}{Q_b} = 1 - \frac{\ln(\theta_0)}{\theta_0 - 1} \qquad (5)$$

The heat content of the hot copper body is $Q_b = C_v(T_o - T_r)$=31.57 kJ/kg. The Carnot work potential $W_G$=3.71 kJ/kg, or Carnot efficiency $\eta_C = W_G/Q_b$=11.8% of the heat content (corresponds to a Carnot average source temperature of 61.3 °C versus arithmetic average of (104+22)/2=63 °C). Note that a coper body used in experiment (1) was a 1 cm³ cube 9-gram mass.

The minimum possible self-sub-cooled body temperature $T_{min}$, corresponding to $\theta_{min}$ and $S_{min}$ (see Fig. 2), could be calculated by equating generated and stored work $W_G$, while cooling the body, Stage 1(G), from initial $T_0$ temperature to $T_r$, ambient room temperature, with $W_C$, sub-cooling work, Stage 2(C), utilized after storage by the cooling device from ambient $T_r$ to $T_{min}$ temperature, including $\varepsilon_G$, $\varepsilon_S$, $\varepsilon_C$, the irreversible efficiencies of the work-generation (G), work-storage-and-retrieval (S), and work-cooling devices (C), respectively, i.e., in dimensionless form:

$$\varepsilon_G \varepsilon_S \int_{\theta_0}^{\theta_r} \left(1 - \frac{1}{\theta}\right) d\theta = \varepsilon_C^{-1} \int_{\theta_r}^{\theta_{min}} \left(\frac{1}{\theta} - 1\right) d\theta \qquad (6)$$

After integration and substitution of limits [note that $\theta_r = 1$ and $\ln(\theta_r) = 0$], the following implicit equation is obtained and solved numerically for $\theta_{min}$ and $T_{min} = \theta_{min} T_r$, for a number of different parameters, see Table 1.

$$F(\theta_{min}) = \varepsilon_G \varepsilon_S \varepsilon_C [\theta_0 - 1 - \ln(\theta_0)] + 1 - \theta_{min} + \ln(\theta_{min}) = 0 \qquad (7)$$

For ideal, reversible processes, $\varepsilon_G = \varepsilon_S = \varepsilon_C = 1$, the above implicit equation reduces to:

$$\theta_0 - \theta_{min} - ln(\theta_0) + ln(\theta_{min}) = 0 \tag{8}$$

Which is equivalent to $\int_{S_r}^{S_0}(T - T_r)dS = \int_{S_{min}}^{S_r}(T_r - T)dS$ or $\int_{T_r}^{T_0}\left(1 - \frac{T_r}{T}\right)dT = \int_{T_{min}}^{T_r}\left(\frac{T_r}{T} - 1\right)dT$, or after integration, the explicit equation reduces to $T_0/T_{min} = exp[(T_0 - T_{min})/T_r]$, which is the same as Eq. (5) in (1).

Note that subcooling of a body at 104 °C initial temperature below 22 °C ambient room temperature, during experimental testing (1), was only about 2 K=2 °C, which corresponds to a very low irreversible efficiency of about 0.001=0.1% of maximum reversible efficiency for which maximum possible subcooling of 69 K=69 °C, i.e., minimum body temperature of -47 °C, would be achieved, see Table 1. A superconducting inductor coil was probably used to increase storage efficiency and provide passive, inertial current switching. Use of a storage electrical capacitor would be less efficient than superconducting inductor and would require manual current switching or to be provided by some automatic current-control electronics. Note also that for self-subcooling of a body, the results as modelled by Eq. (7), depend on the relevant temperature levels and irreversible efficiencies only. If a heat engine or refrigerator irreversible efficiency is 60% each and Peltier generator or cooler 10% each (six times less efficient than heat engine), and 50% capacitive storage-retrieval efficiency (about 75% each), resulting in about 0.5% total irreversible efficiency (0.1*0.1*0.75*0.75=0.005) or about 5 °C subcooling or 16.7 °C minimum body temperature. Even with ideal storage-and-retrieval (superconducting coil), the most efficient Peltier generation-and-cooling will result in about 1% efficiency and 7 °C subcooling (see Table 1). However, with use of classical, efficient heat engine and refrigeration devices with mechanical storage and retrieval (elevating and suspending a weight) the efficiency could be much higher, possibly up to 30% (0.6*0.6*0.9*0.9=0.3) achieving about 40 °C subcooling, i.e., -17.4 °C minimum body temperature. However, since in actual experiments, even with superconductive coil, the subcooling of about 2 °C was achieved, implying only 0.1% over-all irreversible efficiency, probably due to Peltier TE element used and rather low, about 40 °C temperature difference between the variable hot body (Carnot average about 61.3 °C) and 22 °C ambient temperature. Note that the ideal Carnot efficiency was only 11.8% (Eq. 5). That justify almost negligible EM energy stored in inductor compared to the Peltier cooling capacity, and also justify a very poor performance of only about 2 °C subcooling below ambient temperature.

**TABLE 1.** Hot body self-sub-cooling minimum temperature for different irreversible efficiencies of thermal generator, storage-retrieval and cooling devices. The ideal Carnot efficiency, $\eta_C = W_G/Q_b = 11.8\%$, Eq. (5), $T_0$=377K=104 °C and $T_r$=295K=22 °C.

| $\varepsilon=\varepsilon_G \cdot \varepsilon_S \cdot \varepsilon_C$ [1] Eq.(7) | $\theta_{min}$ [1] Eq.(7) | $T_{min}$ [°C] $T_{min}=T_r\theta_{min}$ | $T_{min}-T_r$ [°C] subcooling |
|---|---|---|---|
| 100% | 0.766 | -47.15 | -69.15 |
| 90.0% | 0.777 | -43.90 | -65.90 |
| 85.0% | 0.782 | -42.19 | -64.19 |
| 75.0% | 0.795 | -38.60 | -60.60 |
| 50.0% | 0.830 | -28.18 | -50.18 |
| 30.0% | 0.866 | -17.41 | -39.41 |
| 15.0% | 0.904 | -6.26 | -28.26 |
| 10.0% | 0.921 | -1.22 | -23.22 |
| 5.0% | 0.944 | 5.45 | -16.55 |
| 2.0% | 0.964 | 11.46 | -10.54 |
| 1.0% | 0.975 | 14.52 | -7.48 |
| 0.5% | 0.982 | 16.70 | -5.30 |
| 0.2% | 0.989 | 18.64 | -3.36 |
| 0.1% | 0.992 | 19.62 | -2.38 |

If the sub-cooling process is not stopped after Stage 2(C) on Fig. 2, the Peltier element will perform Stage 3(G) as a generator (G), cycle R-M-N-R, storing EM energy in inductor, due to heat transfer from the ambient to the subcooled body until it reaches the ambient temperature. When ambient temperature is reached the EM energy stored in the inductor will be passively retrieved in reverse and drive Peltier element as a heat pump (H), cycle R-P-O-R, continuing heating the body above the ambient until all stored energy is used and maximum body temperature is reached, Stage 4(H), thus completing the whole cycle (Stages 1-2-3-4). Actually, in experiment (1) the copper body was heated from ambient temperature to initial temperature $T_0$ which is equivalent to Stage 4(H) heat-pump heating, and with the same $W_G$ work-potential stored as body heat, from external source relative to the original ambient equilibrium. If all devices are ideal, the cycle (Stages 1-2-3-4) will continue oscillating indefinitely (compare with Fig. 5). However due to irreversibilities of different kinds, the stored electrical work will be dissipating to heat, and its magnitude will be diminishing after every stage causing the temperature oscillations to be diminishing and ultimately approaching the equilibrium ambient temperature when all stored electrical energy is dissipated.

**Comments on "*Relations to the second law of thermodynamics* (1)"**

The authors (1) *a priori* assumed that the Second law is valid for a perfect Peltier cooler ("the terms related to the Peltier element cancel out,"), and then they accounted only for its obvious irreversibilities (electrical resistance and thermal conduction leak). Of course, the self-serving analysis of *a priori* assuming the final outcome has resulted to expected, "the entropy of the whole system monotonically increases over time."

One may wonder why the authors wanted "to inevitably call for a further discussion in view of the second law of thermodynamics," and why they have not realized after the analysis that there is nothing magic in their so-called new discovery of otherwise a well-known Peltier cooling process.

The authors completely overlooked the well-known phenomena, have not reported (probably have not measured) the local interface-temperatures ($T_{bi}, T_{bP}, T_{rP}, T_{ri}$, see Fig. 1), nor accounted for related and unavoidable entropy generation at the interfaces, regardless how small the temperature differences were (under the circumstances), but significant concept-wise, without which the heat transfer and cooling would not be possible, see Fig. 1 and more detailed analysis presented here.

The authors state "that it [their innovative process (1)] still fully complies with the second law of thermodynamics in the sense that the entropy of the whole system monotonically increases over time, albeit heat is temporarily flowing from cold to hot." They further state that, "The proof that these processes do not violate this fundamental law is surprisingly simple. The total rate of entropy production is $\dot{S}_{tot} = \dot{S}_b + \dot{S}_r = \dot{Q}_b/T_b + \dot{Q}_r/T_r$ ."

However, the above entropy correlation (available in many standard references) only appears simple, but all unavoidable complexities of actual, dissipative processes with entropy generation are implicit in evaluating the heat fluxes and the intervention-work required for cooling. For a Peltier cooling process ($T_b < T_r$), see Fig. 1, we have,

$$\dot{Q}_r = \dot{Q}_b + \dot{W} \gtreqless (T_r/T_b)\dot{Q}_b > \dot{Q}_b \tag{9}$$

$$\dot{W} = \dot{W}_{C(ideal)} + \sum_{all} \dot{W}_{loss} = \dot{Q}_b \frac{T_r - T_b}{T_b} + (I^2 R + \dot{Q}_{leak} \frac{T_r - T_b}{T_b} + \sum_{other} \dot{W}_{loss}) \tag{10}$$

Where, the ideal Carnot cooling-work, dissipated electrical work, heat-leak work-loss, and other dissipated work losses in the Peltier cooler, on the RHS, respectively. See further explanations and relevant comments below.

In addition, the authors (1) used ambiguous nomenclature and signs, and most importantly, did not discuss but ignored, among others, the "direct heat flow from cold to hot," the innovating "magic discovery" as claimed by the authors. Furthermore, as if they were not aware of the required subcooling of the Peltier cold-end interface and overheating of the hot-end interface, to provide spontaneous, local heat transfers from hot-to-cold (as it must always be) in order for cooling process to occur (as a matter-of-concept requirement), then to adiabatically transport heat from cold- to hot-interface using cooling intervention-work *W* (classical heat-transfer should not be confused with heat-transport in Carnot-like cooling devices, the two being different concepts, see also Fig. 1 & 3). The authors also did not account for the entropy generation during the local heat transfer from the cooled body to the Peltier cold-end interface, nor from the Peltier hot-end interface, to the surrounding reservoir, both, concept-wise significant for the cooling process, regardless of their magnitudes.

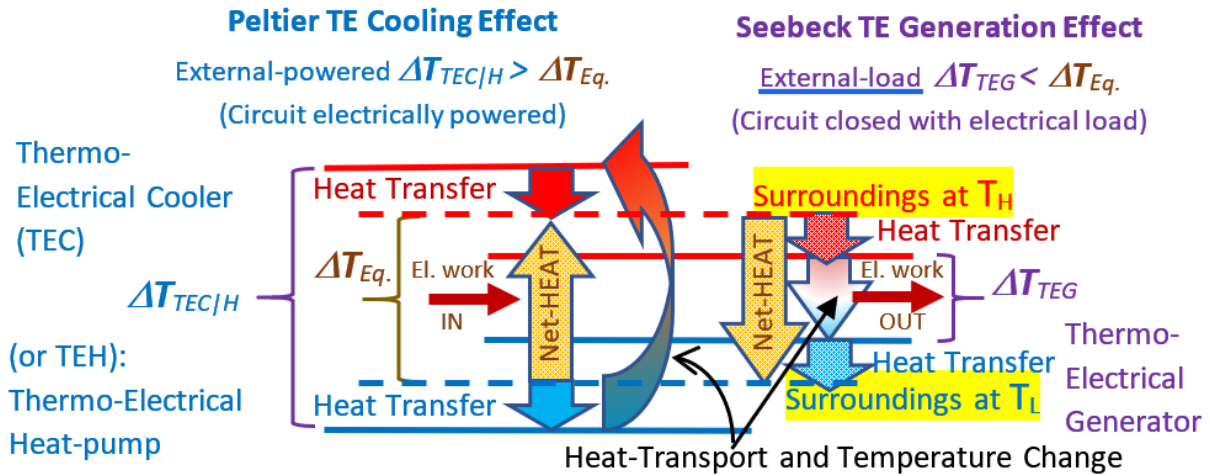

**Fig. 3. Seebeck-Peltier Thermo-Electric Cooler (TEC)**. The equilibrium voltage difference $\Delta V_{Eq.}$ (open circuit TC voltage) is proportional to $\Delta T_{Eq.}$ Temperature of "adiabatically transported thermal energy," by electron current within a TEC element, is transformed (increased or decreased) by in- or out- electrical work (respectively). However, the heat is transferred from high- to low- temperature only at both interfaces of TEC element. Heat cannot actually transfer from cold to hot, only temperature can be transformed (increased or decreased as needed) by work. Heat trans<u>fer</u> (from hot to cold only) should not be confused with trans<u>port</u> of thermal energy (from any to any temperature level).

Additional comments are presented below and elsewhere.

Entropy production rate (aka "entropy generation") is a volumetric, dissipative process and refers to a well-defined control volume (*CV*) of a system (see Fig. 1). In general, the entropy rate balance equation for a system CV includes relevant volumetric quantities and related fluxes through the boundary surface (*A*) of the *CV*, i.e.:

$$\frac{d}{dt}(S_{CV}) = \dot{S}_{gen} + \dot{S}_{in} - \dot{S}_{out}, \text{ therefore, } (\dot{S}_{gen} = \dot{S}_{tot}) = -\dot{S}_{in} + \dot{S}_{out} + \frac{d}{dt}(S_{CV}) \qquad (11)$$

where, $S_{CV} = \int_{CV} s\,dV$, $\dot{S}_{gen} = \int_{CV} \sigma\,dV$, $\dot{S}_{in} = \int_A \dot{s}_{in}\,dA$, and $\dot{S}_{out} = \int_A \dot{s}_{out}\,dA$, are the system entropy, entropy generation rate, and in- and out- entropy boundary fluxes, respectively.

Only for non-transient (stationary, i.e., steady state) processes a system entropy production rate or entropy generation is $\dot{S}_{gen|tot} = \dot{S}_{net\_out} = \dot{S}_{out} - \dot{S}_{in}$ ("*in & out*" with respect to the CV), when properly accounted for the boundary surface *A* of the *CV* of the system (Peltier element with the cold- and hot-end interfaces in this case).

The $\dot{Q}_b = k(T_b - T_r)$ was defined in introduction (1) as heat from a body (*b*) through a system (*k* thermal conductance) and into a surrounding reservoir (*r*), and regrettably the same correlation used later for the Peltier heat-conduction leakage, although the latter is a fraction of the former, see below. In addition, the nomenclature ambiguity and sign inconsistency introduce further confusion. Note that the complete entropy calculations details were not presented in (1). The comments below are presented as a matter-of-concept for the steady state process but equally apply in general and after proper time-integration for a transient process as well.

In general, for cooling devices $|\dot{Q}_b| \neq |\dot{Q}_r|$, but $\dot{Q}_r = \dot{Q}_b + \dot{W}$ is bigger than $\dot{Q}_b$ for the "cooling intervention" work rate $\dot{W} = \dot{W}_C + \sum_{all} \dot{W}_{loss} = \dot{W}_C + (I^2 R + \dot{Q}_{leak}(T_r - T_b)/T_b + \sum_{other} \dot{W}_{loss})$ (the ideal Carnot work, dissipated electrical work, heat-leak work-loss, and other dissipated work losses in the Peltier cooler, on the RHS, respectively). In all cooling devices ($T_b < T_r$), the $\dot{Q}_r \gtreqqless (T_r/T_b)\dot{Q}_b > \dot{Q}_b$, the Clausius inequality, where $T_r/T_b = \dot{Q}_r/\dot{Q}_b$, is the Carnot-ratio in the Clausius equality (note that absolute-value notation is dropped since all quantities here are positive, to avoid any ambiguity).

If $\dot{Q}_b = \dot{Q}_r$, like in direct heat transfer through a wall, then for a Peltier cooler ($T_b < T_r$) without external intervention ($\dot{W} = 0$), the direct heat transfer, if possible, would result in violation of the Second law, thus negating possibility of "direct heat transfer from cold to hot," i.e.:

$$\dot{S}_{tot,dir} = -\dot{S}_{in} + \dot{S}_{out} = \left(-\dot{S}_b + \dot{S}_r\right)_{dir} = -\frac{\dot{Q}_b}{T_b} + \frac{\dot{Q}_r}{T_r} = \dot{Q}_b\left(-\frac{1}{T_b} + \frac{1}{T_r}\right)$$

$$= \frac{-\dot{Q}_b(T_r - T_b)}{T_r T_b} < 0 \qquad (12)$$

Note that this does not apply to an actual Peltier cooler, since $\dot{Q}_b$ heat is not being directly transferred as claimed in (1) (except for a small heat leakage fraction, always from hot to cold, $\dot{Q}_{leak} \cong k(T_r - T_b) \ll \dot{Q}_b$), but adiabatically (ideally isentropically) transported with TEC electron stream, driven by the electrical intervention-work from inductor (or any other electrical source). During the cooling process the electron stream temperature was decreased below $T_b$ to $T_{bP}$ at the body interface and increased above $T_r$ to $T_{rP}$ at the reservoir interface so that actual heat is being always spontaneously transferred from higher to lower temperature locally, while overall transported from cold-end to hot-end of Peltier cooler, like in all other refrigeration devices (adiabatic heat transport is concept-wise different from actual heat transfer), see Fig. 1.

As already stated, the entropy generation analysis (1) is not comprehensive but oversimplified, ambiguous, and incomplete. The authors assumed that the ideal Peltier-inductor system process does not generate any entropy (as it should not violate the Second law) and only included the very common Peltier-cooler dissipative processes, due to its final values of electrical resistance ($R$) and thermal conductance ($k$). Namely, electrical current dissipation, $\dot{S}_{gen,R} \cong (I^2R)\left[\frac{1}{2}\left(\frac{1}{T_b}+\frac{1}{T_r}\right)\right] > 0$ (actually dissipated within the CV but assumed to be divided equally towards the two Peltier ends), and heat conduction leak through, $\dot{Q}_{leak} = k(T_{rP} - T_{bP}) \cong k(T_r - T_b)$, always from hot- to cold-end, $\dot{S}_{leak} \cong \frac{-k(T_b-T_r)}{T_b} + \frac{k(T_b-T_r)}{T_r} \geq 0$, should be zero in ideal Peltier cooler (if the perfect electrical conductor and thermal insulator, $R=k=0$).

Namely, the undesirable heat-conduction leak is through the Peltier interface surface ($A_{PI}$) along its thickness ($L_P$) with effective conductance $k = k_P A_{PI}/L_P$ ($k_P$ being the thermal conductivity of Peltier element, preferably as small as possible), and the related entropy generation, $\dot{S}_{leak} \cong \dot{Q}_{leak}\left(\frac{1}{T_b} - \frac{1}{T_r}\right) = k(T_r - T_b)\left[\left(\frac{1}{T_b} - \frac{1}{T_r}\right)\right] = k(T_r - T_b)^2/(T_r T_b) > 0$.

Note that all dissipative processes of any work-loss to generated-heat (and accompanying entropy generation) are volumetric as opposed to surface fluxes and always in direction opposite to (i.e., on the expense of) the transported cooling load, albeit may be approximated by the average surface quantities, and assigned to be split at the two ends of the cooling device. The total cooling intervention-work $\dot{W}$ must be supplied to provide for ideal, reversible Carnot work $\dot{W}_C$ and any and all additional lost works, $\sum_{all}\dot{W}_{loss}$, dissipated to generated heat ($\dot{W}_{loss} = \dot{Q}_{gen}$), ultimately at the surrounding reservoir temperature $T_r$, i.e.:

$$\left(\dot{S}_{gen} = \dot{S}_{tot}\right) = \frac{\sum_{all}\dot{W}_{loss}}{T_r} = \frac{\dot{Q}_r}{T_r} - \frac{\dot{Q}_b}{T_b} + \frac{d}{dt}(S_{CV}) \geq 0 \qquad (13)$$

The last transient term, *d/dt*, is zero for steady state processes and may be neglected for relatively slow processes. The Peltier heat conduction leak $\dot{Q}_{leak} \cong k(T_r - T_b)$ should not be confused with "direct heat transfer" presented by the authors as a "miracle discovery," nor with the cooling load, the heat flow from the body through the cold-end interface, $\dot{Q}_b = |k_{bP}(T_b - T_{bP})| \neq |k(T_b - T_r)|$, to the surrounding reservoir ($\dot{Q}_r = |k_{rP}(T_{rP} - T_r)| = \dot{Q}_b + \dot{W} > \dot{Q}_b$), see Eq.(9) and Fig. 1.

The "direct heat transfer" (as hypothesized by the authors) is not a real heat-transfer but actually, adiabatic heat transport ($\dot{Q}_{port} = \dot{Q}_b + \dot{W} = \dot{Q}_r$), carried by a flow of a coolant medium (the Peltier electron stream or refrigerant medium in general). The adiabatic heat transport is not associated with entropy and ideally does not generate any entropy. The cooling-load heat ($\dot{Q}_b$), transported by electron stream ($\dot{Q}_{port}$) from the body ($b$) through the Peltier cooler (TEC) into the surrounding reservoir ($\dot{Q}_r$), i.e.,

$$\dot{Q}_{port} = \dot{Q}_b + \dot{W} = \dot{Q}_b + (\dot{W}_C + I^2R + \dot{Q}_{leak} + \Sigma_{other}\dot{Q}_{gen}) = \dot{Q}_r \gtreqqless (T_r/T_b)\dot{Q}_b > \dot{Q}_b \qquad (14)$$

The dissipation losses, $\dot{W}_{loss} = \dot{Q}_{gen}$, sometimes appear entangled and elusive, and could be expressed and grouped in deferent ways, but if properly and consistently accounted for, should always produce the same final result.

**HEAT IS TRANSFERRED FROM HOT TO COLD WITHOUT EXCEPTION**

It appears at first that heat is transferred from cold to hot in refrigerators and other cooling and heat pump devices. However, a careful observation will reveal that objects (like food in a refrigerator) is cooled by transferring heat to even colder evaporator surface inside refrigerator, then compression of refrigerant (with work input) will transport its energy and increase its temperature above the surrounding ambient so that heat will be transferred from hot outside condenser-surface to the ambient-medium at lower temperature than the refregerant. In the complete process, the thermal energy will be transported from cold medium to a hot medium, while actual heat transfers will always be taking place from hotter to colder temperature, without any exception, see Fig. 1, 3 & 4. It is important to differentiate between the actual, direct heat transfer (always from hot to cold) from transport of thermal energy (from any-to-any temperature level). On Figs. 1 & 3 a Seebeck-Peltier thermo-electric device, as used in (1) is depicted, while on Fig. 4 an ideal thermal transformer is represented, consisting of reversible power- and refrigeration Carnot cycles (8). More details are provided on the Figures' captions and in the Discussion section.

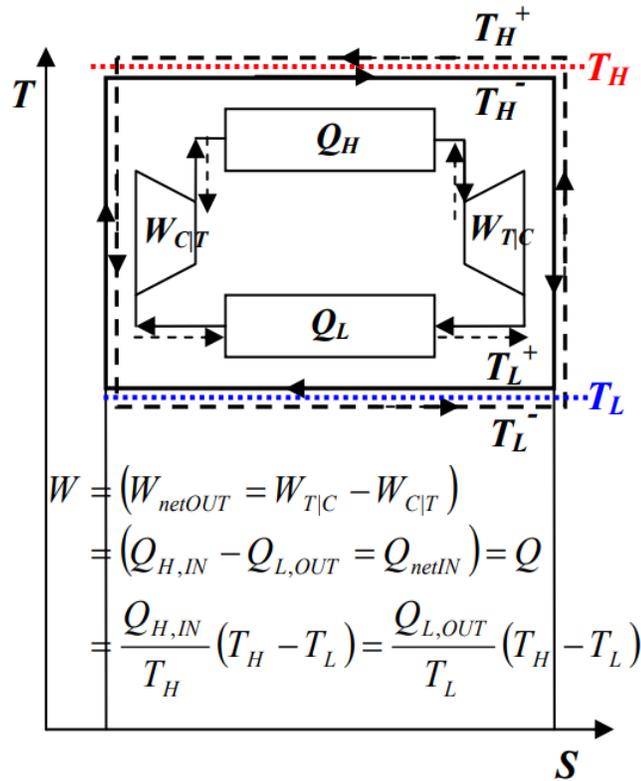

**Fig. 4. Ideal Thermal Transformers - Reversible Carnot cycles** (8): Heat is net-transferred from-any-to-any temperature levels. Thermal reservoirs with high ($T_H$) and low ($T_L$) temperatures are presented as dotted lines. Heat-engine cycle presented by solid lines, and Refrigeration cycle presented with dashed lines (reversed directions). With isentropic compression and expansion the temperature of the working cycle agent is adjusted infinitesimally below or above the reservoirs' temperatures to either transfer heat from high to low temperature and extract turbine-compressor net-work out (solid lines), or using net-work in from outside to transport heat from low to high temperature (dashed lines, either cooling or heat pump processes). Ideal reversible cycles provide, in limit, reversible heat transfer from higher to lower temperature and extraction of maximum thermal work potential, and reversible heat transfer from lower to higher temperature with expenditure of minimum possible work, thus establishing reversible equivalency and ideal thermal transformer. Note that actual heat transfer is always from at least infinitesimally higher to lower temperature, resulting in net-heat transport from any to any temperature.

**THERMAL TRANSFORMER AND TEMPERATURE OSCILLATORS**

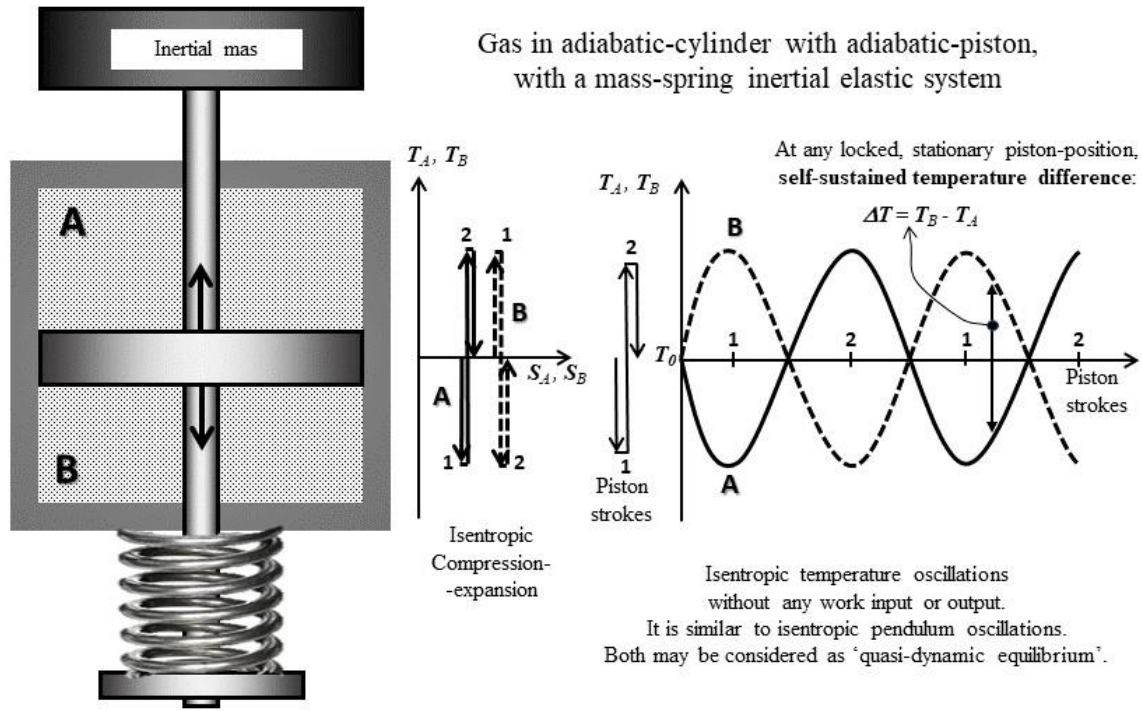

**Fig. 5. Temperature Oscillator - Gas in adiabatic piston-cylinder system isentropically produces perpetual temperature oscillations or temperature difference.** If the depicted piston, from isothermal center-position, is compressing an ideal gas in partition B, thus isentropically increasing the gas temperature, then gas will expand in partition A and isentropically decrease its temperature, without any heat transfer. If displaced piston within the cylinder with ideal, inertial mass and elastic spring system is left free, it will perpetually oscillate, but without any perpetual work generation, similarly to an ideal pendulum oscillations, thus demonstrating a thermal 'dynamic quasi-equilibrium' with perpetual temperature oscillations (but not entropy oscillations); or, at any locked, stationary piston position, a self-sustained 'structural quasi-equilibrium' will establish with perpetual temperature difference, without violating the Second law.

An isentropic, macro temperature transformer-oscillator, capable of producing perpetual temperature oscillations or perpetual temperature difference, similar to underdamped temperature oscillations on Fig. 1D & 3B (1), is depicted on Fig. 5. The both temperature transformer-oscillators, adiabatic piston-cylinder on Fig. 5 and a "thermal inductor" in (1) are non-thermal by nature but driven by stored work and related mechanical and EM inertia, respectively. If a piston in the cylinder with ideal, inertial mass and elastic spring system, at isothermal center-position, is compressing an ideal gas in partition B, thus isentropically (non-thermally) increasing the gas temperature, then gas will expand in partition A and isentropically decrease its temperature (Fig. 5). If the displaced piston is left free, it will 'perpetually oscillate', but without any perpetual work generation, similarly to an ideal pendulum oscillation, thus demonstrating a thermal 'dynamic quasi-equilibrium' with perpetual temperature oscillations. At any locked, stationary piston position, a self-sustained 'structural equilibrium' will establish with perpetual temperature difference, as depicted on Fig. 5, without violating the Second Law (entropy is conserved, no entropy destruction). The initial compression work may be obtained back if original equilibrium is re-established in a transient reversible process. However, such non-uniform temperature cannot be utilized for perpetual work generation without perpetual work consumption from elsewhere.

Similarly, the perpetual dynamic oscillations or fluctuations at micro or macro scales may not be utilized for perpetual work generation. It is shown on Fig. 5 that even macro fluctuations, after a transient start-up process, may be set into perpetual oscillatory motion without any external perpetual work input or output. Therefore, only limited transient, but not a perpetual work, may be obtained from such oscillating or fluctuating systems. As depicted on Fig. 5, an ideal adiabatic piston-cylinder with gas, may demonstrate a thermal-mechanical, structural equilibrium with non-uniform temperature or perpetual temperature oscillations, without violating the Second Law of thermodynamics.

Fluctuating phenomena in perfect equilibrium are reversible, thus isentropic, so that any reduction in entropy reported in the literature (whatever that means), may be due to 'improvised' and incomplete entropy definitions at micro- and sub-micro scales, approximate accounting for thermal and/or neglecting diverse displacement contributions to entropy (for example, during isothermal free expansion of an ideal gas, entropy is generated due to elastic-field volume-displacement, $\delta S_{gen}=dS=R_{gas}dV/V$, and similarly it could be due to subtle and illusive micro- and sub-micro forced-field 'displacements', including particle 'correlation' and/or quantum entanglement in respective force fields, etc.).

In summary, there are three types of self-sustained macro equilibriums (stable and perpetual, with balanced macro-forces, including inertial forces, and net-zero mass-energy fluxes): (1) *homogeneous equilibrium* with uniform-properties, e.g., thermo-mechanical equilibrium within simple fluids and solids, as in classical thermodynamics; (2) *dynamic equilibrium* (a.k.a. dynamic quasi-equilibrium) with spontaneous and perpetual self-sustained motion, like ideal thermal oscillations on Fig. 5, ideal pendulum oscillations, orbiting electrons around a nucleus, thermal-molecular random-motion, etc.; and (3) *structural equilibrium* (a.k.a. "boundary constrained' quasi-equilibrium) with heterogeneous and non-uniform properties, like mechanical, thermal, electrical and chemical potentials, i.e.: (*i*) hydrostatic pressure in gravity field, (*ii*) inflated air-balloon; (*iii*) hot medium in an adiabatic thermos flask; (*iv*) electro-chemical cell; (*v*) fuel with air (ready for combustion), etc.

Dynamic and structural quasi-equilibriums with non-uniform properties, as compared with classical 'homogeneous equilibrium', are elusive and may be construed as non-equilibriums, if they are re-structured and in a transient process some useful work is obtained, to allude to violation of the Second Law.

However, such work-potential is limited to one stored within such a structure (regardless how small or large), and cannot be utilized as a perpetual (stationary or cyclic) *Perpetual Motion Machine of the Second-kind* (PMM2 device), or Maxwell's Demon (MD) or similar, to continuously generate useful work from within an equilibrium. There may be some similarity with the "hot body self-sub-cooling" (1), where initial non-uniform properties have been used for transient subcooling or underdamped oscillations, to allude to some miraculous phenomena which they are not.

Most of the fallacies of diverse PMM2, MDs and Second Law challenges, or even hypothetical violations, are related to elusive dynamic or structural quasi-equilibriums when transient work potentials are hypothesized to be perpetual. Engaging any engine or engine-like structure to produce work may run only transiently until existing physical and electro-chemical work-potentials are exhausted and another equilibrium is re-established with such a new device or structure. Testing such hypothetical devices, to produce perpetual work from within surrounding equilibrium, would be straight-forward and not difficult, but somehow overlooked by the Second Law challengers. However, a PMM2-Second Law violation has never been confirmed possible, but to the contrary. Therefore, speculating that such engines or devices will work perpetually and violate the Second Law is only an illusory imagination and physically impossible.

**DISCUSSIONS AND CONCLUSIONS**

The authors (1) acknowledge compliance with the Second law of thermodynamics by calculating simplified entropy generation but claim that other phenomena are unexpected and miraculous, i.e.,: "*there have been exciting technical developments during the past few years, which appear to be at odds with popular interpretations of the laws*." The authors further state, "*We show that it [flow of heat from cold to hot] can also act in a passive way without any external or internally hidden source of power*." In a follow up article (3), Schilling, the leading author of (1), further state, *"Such an intervention would do the work to force heat to flow from cold to hot and, ultimately, always requires an external energy source… the work is more significant than a mere "proof-of-principle" study."* Regrettably, the following statements are fundamentally inaccurate: "*A subtle but important point to mention here is the fact the heat was directly flowing from cold to hot in these processes, without ever being converted into any other form of energy. It turned out that the magnetic energy stored in the coil is negligible, and therefore the heat must have flown directly from cold to hot.*" However, direct heat flow from cold to hot will destroy entropy and thus violate the Second Law. Regrettably, Schilling *et al* (1) had not measured the temperature difference between the end-plates of the Peltier element and the surrounding mediums, to verify that direct heat was always flowing from hot to cold (regardless how small temperature difference) and not otherwise, see Fig. 1 & 3. In fact, magnetic energy stored in the coil inductor provides the Carnot work for cooling process and is thermodynamically very small (may appear negligible) for the very small sub-cooling temperature difference of about 2 °C (as in experiment), i.e., $W_c/Q_b=(T_b-T_r)/T_b=2/293=0.007=0.7\%$ (see also related discussions elsewhere). Therefore, the heat was actually always flowing from higher to lower temperature, from cold body to even colder cold-interface of the Peltier element, and then the temperature of the opposite side of the Peltier element was transformed by electrical energy stored in the inductor (in a non-thermal adiabatic process) to even higher temperature than the room ambient, so that heat was flowing again from higher to lower temperature (see Fig. 1 & 3). Regrettably again, the authors (1) have not measured the Peltier element cold- and hot end-interface temperatures that were somewhat colder and hotter from the cold body and ambient temperatures, respectively, during the subcooling process for example (regardless how good thermal contacts with Peltier were, they could not have been perfect). Important but subtle point is not to confuse heat with temperature, and not to confuse actual, direct heat transfer (Fourier "flowing" heat) with overall, "net-transport of thermal energy by work" from cool to hot ambients.

**Summary of the fundamentally false claims in (1):**

1. "*Heat flow from cold to hot without external intervention, and without any source of power*," is false claim since the sub-cooling the body below ambient temperature by a Peltier element (Fig. 2), Stage 2(C) process, was driven by the "external intervention" using electrical work retrieved from stored EM energy in the inductor (external to the Peltier element), stored in the prior Peltier generation process, Stage 1(G), while cooling the body to ambient-temperature. There is nothing miraculous in Stage 1(G) power generation process, and the EM energy stored is indeed an "external intervention" by all standards of thermodynamics, when used in Stage 2(C) Peltier cooling process, see Fig. 2.

2. "*Direct heat transfer from cold to hot in the Peltier TEC element*," is false claim since the heat cannot actually, directly transfer from cold to hot (would be destruction of entropy and violation of the Second Law). It only can be transported with material medium (a refrigerant or electron stream) while temperature of the medium is transformed by work, increased or decreased to drive heat transfer as desired (as in all thermal-power, cooling or heat pump devices). Direct heat transfer (from hot to cold only) should not be confused with transport of thermal energy (from any to any temperature level). The heat is actually transferred from the subcooled body to the even lower temperature of the Peltier cold end-plate, and after being adiabatically transported with temperature increase by electron stream (similar to a refrigerant medium transport and compression), the heat is then actually transferred from higher temperature of the Peltier hot end-plate to the lower temperature of the ambient air, see Fig. 1 & 3. There are no reported temperature measurements of the Peltier cold and hot end-plates (1), to compare with hot body and ambient temperatures, respectively (regardless how small temperature difference was). The authors might be deceived by negligibly small inductor energy (smaller than 0.01% of transported heat, but due to low Carnot's and extremely low experimental irreversible efficiencies) and mistakenly thought "*that some miraculous flow of heat from cold to hot is taking place in Peltier-inductor circuit without external or internal intervention.*"

3. *The "thermal inertia"* is false claim since the underdamped oscillations were driven by external, non-thermal but adiabatic "inductor magnetic inertia," driving temperature oscillations in Peltier element, followed by dissipative, non-inertial *per se*, heat transfer oscillations. Only if the work potential is not fully dissipated (as is during actual, direct heat transfer), but if it is extracted and stored to provide adiabatic, non-thermal inertia to force temperature oscillations followed by non-inertial heat transfer oscillations or only sub-cooling as desired (Stage 2 end of subcooling).

4. "*Use of inductor is essential*," is false claim since, although the high efficiency of the superconducting coil was beneficial, but the thermal oscillations are not essential, actually not desirable at all, since the first half of the cycle (Stages 1 & 2 on Fig. 2) provide the maximum sub-cooling, the oscillations being non-essential (actually adverse) by-product of the inductor-Peltier circuit oscillations. The same effect could have been achieved with any work-energy storage if efficient. Supper-conductive inductor provides high efficiency and "magnetic inertia" (not "thermal inertia"), also passive reversal of stored energy (capacitor switching would be required) to run Peltier element as a cooling device. The sub-cooling could be achieved with other work-storage devices, therefore the inductor *per se* is not essential, see Fig. 2 that demonstrate subcooling processes in general.

The authors (1) demonstrate compliance with the Second Law "*in the sense that the entropy of the whole system monotonically increases over time, albeit heat is temporarily flowing from cold to hot*," but believe that the novel device, "*to our knowledge, never been considered in the literature*,"  is undergoing miraculous processes of "*Heat flowing from cold to hot without external intervention by using a "thermal inductor" … and presents "a new discovery beyond the proof of concept.*" As if Schilling is challenging the Second law of thermodynamics (1, 3): "*Nonetheless, the experiments challenge to some extent our ordinary perception of the flow of heat, which may have been misguided by certain shortened versions of the second law of thermodynamics in some textbooks.*" As if the Second Law is violated locally by "direct heat flow from cold to hot," the main new discovery by Schilling et all (1), but overall in compliance with the Second Law, as if the "monotonic entropy increase over time" was magically compensated somehow? Some "ordinary perceptions of the flow of heat" are due to lack of subtle comprehension of the Second Law and its misinterpretations, and are not "misguided by certain shorten versions of the second law." There are no "longer and shorter versions" of the Second Law: all versions of the Second law are equivalent, since all could be correlated with well-known "reversible equivalency." The Second Law is omnipresent at all space and time scales, and valid for closed and open systems, for non-animate and animate systems, without any exceptions; that is that non-equilibrium work potential is in part and ultimately in whole, irreversibly converted to heat, thus always, irreversibly producing (i.e., generating) entropy, since there is no way to destroy entropy, the latter not to be confused with entropy decrease due to its transfer elsewhere (9).

If the authors (1) believed that a magical "direct heat transfer from cold to hot" is due to "thermal inductor, a novel thermal element, able to drive the temperature difference between two massive bodies to change sign by imposing a certain *thermal inertia* on the oscillatory flow of heat [they originally designed for precise heat capacity measurements]," then they are fundamentally mistaken. Actually, the "thermal inertia" *per se* is physically impossible (heat cannot spontaneously transfer from cold-to-hot to "undershoot and/or overshoot" the existing thermal source or sink (as it could be done with cooling intervention-work), but the oscillatory heat transfer is driven by electro-magnetic inertia of the Peltier-inductor system with storage-and-retrieval of the electrical work (not dissipative heat) that drives adiabatically (not thermally) the oscillating temperature change, that in turn drive non-inertial heat transfer until all stored-and-retrieved electrical energy dissipate to heat.

It could be easily confirmed experimentally that *thermal inductor* oscillations are concept-wise irrelevant, by driving the same Peltier cooler with a steady electrical source equivalent to the inductor level. Afterall, the maximum cooling effect is achieved in the first half cycle while the proceeding oscillations are not necessary but are adverse and irrelevant.

In summary, publication (1) was only an interesting and creative "show-and-tell" innovative application, although impractical, but not miraculous nor unusual and "beyond the-proof-of-concept" as presented. The self-cooling of a hot body (with stored work-potential within) could have been achieved by any work generating device, to utilize the stored work-potential, and store it within any suitable device (superconductive inductor was beneficial but not essential as claimed), and such stored work (as external intervention, "external" being relative) used subsequently in any refrigeration device to sub-cool the body. It is hoped that this treatise will also help demystify some recent challenges of the Second Law of thermodynamics and promote constructive future debates.